\documentclass[twocolumn]{aastex62}

\usepackage{graphicx}
\usepackage{amsmath}
\usepackage{amssymb}
\usepackage{enumitem}

\usepackage{natbib}

\def \lya  {Ly$\alpha$}

\def \ly5  {Ly-5}
\def \ly6  {Ly-6}
\def \ly7  {Ly-7}
\def \nhi  {$N_{\rm HI}$}

\newcommand{\nqsos}{727}
\newcommand{\nspectra}{872}
\newcommand{\ntotspectra}{2753}

\newcommand{\hi}{\ion{H}{1}}

\newcommand{\civ}{\ion{C}{4}}

\newcommand{\ovi}{\ion{O}{6}}

\newcommand{\mgii}{\ion{Mg}{2}}

\shortauthors{O'Meara et al.}
\shorttitle{KODIAQ DR3}

\begin{document}

\title{The Third Data Release of the KODIAQ Survey}

\author[0000-0002-7893-1054]{John M. O'Meara}
\affiliation{W.M. Keck Observatory 65-1120 Mamalahoa Highway Kamuela, HI 96743}

\author[0000-0001-9158-0829]{Nicolas Lehner}
\affiliation{Department of Physics, University of Notre Dame, Notre Dame, IN 46556}

\author[0000-0002-2591-3792]{J. Christopher Howk}
\affiliation{Department of Physics, University of Notre Dame, Notre Dame, IN 46556}

\author[0000-0002-7738-6875]{J. Xavier Prochaska}
\affiliation{UCO/Lick Observatory, Department of Astronomy \& Astrophysics, University of Califorinia Santa Cruz, 1156 High Street, Santa Cruz, CA 95064}

\begin{abstract}
We present and make publicly available the third data release (DR3) of the Keck Observatory Database of Ionized Absorption toward Quasars (KODIAQ) survey.  KODIAQ DR3 consists of a fully-reduced sample of \nqsos\ quasars at $0.1 < z_{\rm em} <  6.4$ observed with ESI at moderate resolution ($4000 \le R \le 10000$).  DR3 contains \nspectra\ spectra available in flux calibrated form, representing a sum total exposure time of $\sim 2.8$ megaseconds. These co-added spectra arise from a total of \ntotspectra\ individual exposures of quasars taken from the Keck Observatory Archive (KOA) in raw form and  uniformly processed using a data reduction package made available through the XIDL distribution.  DR3 is publicly available to the community, housed as a higher level science product at the  KOA and in the \textit{igmspec} database.  
\end{abstract}


\section{Introduction}\label{intro}
Studies of absorption lines towards quasars have impacted a wide range of fields within astrophysics and cosmology.  From studies of individual systems to determine the primordial D/H ratio (e.g., \citealt{cooke18}), to studies of tens of high resolution, high signal-to-noise spectra to constrain the temperature-density relation in the IGM \citep{hiss19}, to a determination of the mean free path of ionizing radiation from a stack of thousands of lines of sight \citep{prochaska09}, quasar lines of sight have become one of the most valuable tools available to understand the universe at cosmological distances.  

Large quasar and galaxy surveys such as SDSS \citep{sdss20} provide spectra in bulk in a ready-for-analysis state.  By contrast, quasar spectra from larger ground-based facilities are often not available to the community in a science-ready format, if at all, and with little information on the provenance of the data.  To help remedy the situation, a number of groups have provided the community with pipeline-reduced data from large aperture telescopes (e.g., \citealt{murphy18}).  As a part of these endeavours, a primary goal of the Keck Observatory Database of Ionized Absorption toward Quasars (KODIAQ, \citealt{lehner14}) is to uniformly reduce and make publicly available spectra of quasars obtained with instruments on the Keck telescopes over the last two-plus decades.  In the first \citep{omeara15} and second  \citep{omeara17} data releases (DR1 and DR2, respectively), KODIAQ has provided spectra from Keck's high-resolution echelle spectrograph HIRES.  The DR2 sample encompasses DR1 and adds additional spectra, bringing the sample to a total of 300 unique quasar sightlines.  In this work, we significantly increase the number of quasars with the third data release (DR3) from KODIAQ, this time adding \nqsos quasar spectra obtained with the Echellette Sepctrograph and Imager (ESI, \citealt{sheinis02}) instrument on the Keck~II telescope.  Although lower resolution than HIRES, ESI spectra are of significant resolution to both serve statistical studies of the Lyman-$\alpha$ forest as well as metal line absorption, along with studies of individual absorbers such as damped Lyman Alpha (DLA) systems (e.g. \citealt{rafelski12}).

\begin{figure*}[ht]
\epsscale{0.85} 
\plotone{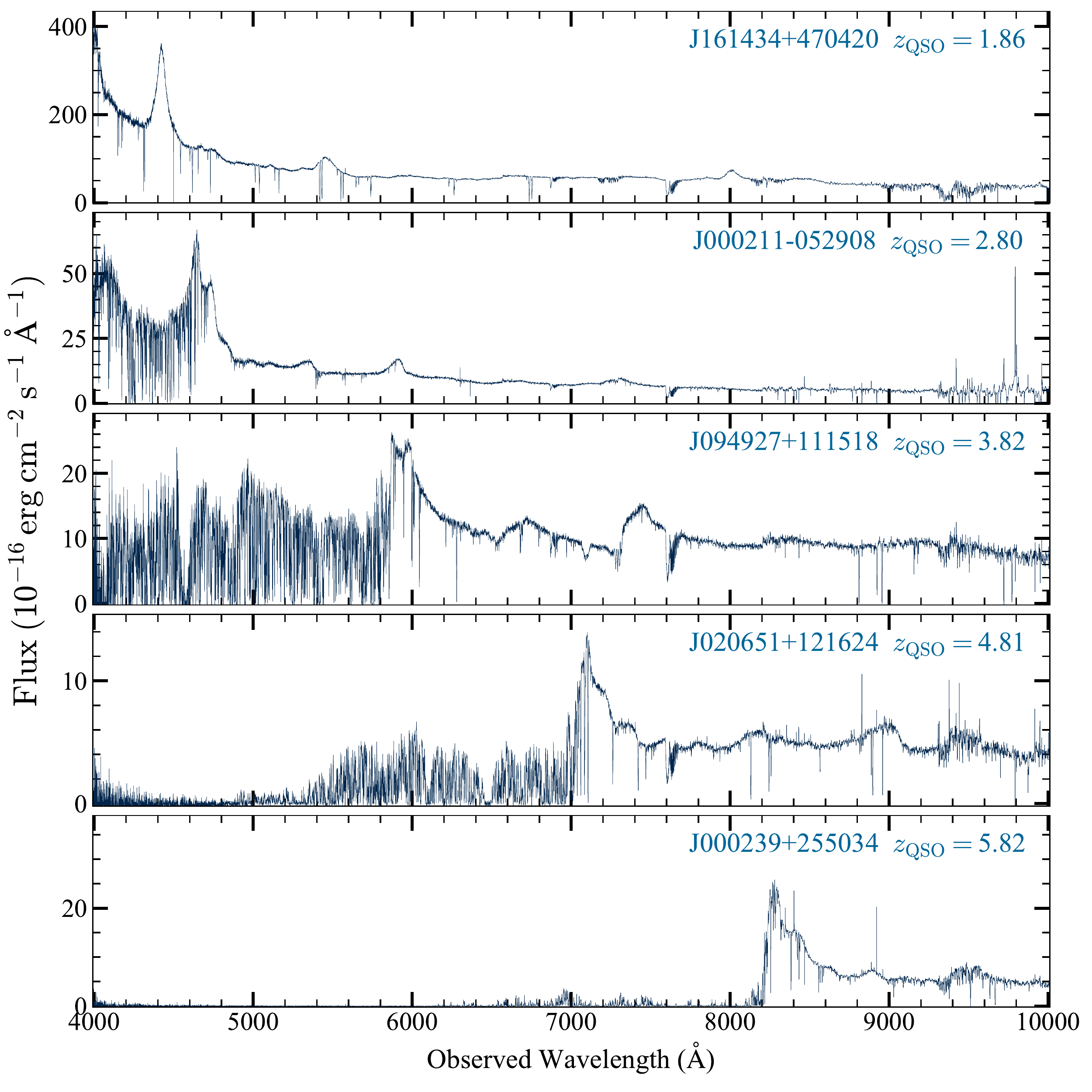}  
\caption{Examples of QSO spectra from the DR3 with increasing QSO redshift from the top to bottom panels. \label{fig_example}}
\end{figure*}

\section{The Data}\label{data}
The ESI data in DR3 were obtained by individual PIs between December, 1999 (shortly after instrument commissioning) and January, 2015.  ESI is an echellette system, with ten echelle orders spanning the range $3930 < \lambda < 10930$ \AA. Cosmetic defects in the ESI CCD impact some pixels in the wavelength regions $4460 < \lambda < 4565$ \AA.  Depending on seeing conditions, slit width, and desired wavelength, ESI has a resolution of $4000 < R < 10000$. Observations in DR3 are made using the full range of ESI slit widths, ranging between 0.3 and 6.0 arcseconds, with the majority of observations made with the 0.5, 0.75, and 1.0 arcsecond slits.  The observations are summarized in Table~\ref{obs_summary}.  The coordinates and quasar emission redshifts presented in Table~\ref{obs_summary} were obtained by passing the coordinates from the raw data file first through the Simbad\footnote{http://simbad.u-strasbg.fr/simbad/} databases and then, if not found, the NED\footnote{https://ned.ipac.caltech.edu} database, selecting the quasar closest on the sky.  If an object was not found within thirty arcseconds of a quasar in either database, the coordinates were taken directly form the data header.  We caution that in the case of close quasar pairs, the coordinates, chosen to correspond to the nearest object, may be inaccurate and refer to the other pair member.  

\subsection{Data Reduction}\label{redux}
Once the data were downloaded and reviewed for calibrations, they were uniformly reduced using the ESIRedux\footnote{https://www2.keck.hawaii.edu/inst/esi/ESIRedux/} code.  ESIRedux is a suite of IDL routines within the XIDL\footnote{https://github.com/profxj/xidl} package of astronomical utilities.  ESIRedux is functionally very similar to HIRedux, which is described in \citet{omeara15}, but with a few important differences which we outline below.  The workflow of ESIredux is as follows:  First, the raw data is grouped into setups.  Unlike for HIRES, ESI is essentially fixed in wavelength coverage and echelle order placement on the detector, so setups are governed only by slit width and detector binning.  Next, for each setup, a bias level and gain is determined, and a flat field image is constructed from a combination of internal lamps, dome flats, and if present, twilight sky flats.  The flat field images serve the traditional role of removing pixel-to-pixel variations, but also provide the location of the slit edges for each echelle order.  Twilight flats are used primarily in the bluest orders, where the internal and dome flats have lower flux.  A two-dimensional wavelength image for the detector is then created using a combination of CuAr, HgNe, and Xe lamps.  The wavelength solutions are compared to archived databases, and the residuals to the fit are most commonly found to be sub-pixel.  Cosmic ray rejection is then performed using standard frame comparison techniques.  As with HIRedux, sky subtraction and object definition and extraction is performed using the methods outlined in \citet{bochanski09} on an order-by-order basis, resulting in a spectrum of each order with counts as a function of wavelength.  Finally, for objects with multiple exposures, the individual orders are weighted mean combined.

A key difference between the ESI data in KODIAQ DR3 and the HIRES data from DR2 is that the data in DR3 have been flux calibrated instead of continuum normalized.  For expediency, all exposures in DR3 were flux calibrated using a single archival sensitivity function derived from a median of sensitivity functions from multiple flux standard stars. Examples of DR3 ESI QSO spectra are shown in Fig.~\ref{fig_example}. 

\subsubsection{Issues of note}
A number of issues should be noted before using data from DR3 for precision analysis.  First, the reddest echelle order, with a central wavelength of approximately $\lambda = 9800$ \AA\ was often poorly extracted.  In these cases, that order has been rejected, and the data trimmed.  The rightmost column in Table~\ref{obs_summary} indicates the wavelength range of the spectra in DR3.  Second, the detector defects at $\lambda \approx 4500$ \AA\ are present in the reduced data, and caution should be used near those wavelengths.  Finally, some spectra suffer discontinuities at the regions in wavelength where the echelle orders overlap.  These discontinuities are the result of order-to-order differences in flux calibration.  Users that intend an analysis of the intrinsic quasar spectra should first inspect the data for these occurrences.  We note that all intermediate data reduction steps for the data in DR3 are provided in the Keck Observatory Archive (KOA),\footnote{https://koa.ipac.caltech.edu/Datasets/KODIAQ/} so that users may adjust specific steps, such as flux calibration, to match their needs.  Finally, Table~\ref{obs_summary} shows that a number of quasars were observed on multiple dates, and often by multiple PIs.  We did not attempt to combine the data across these observations, given frequent differences in slit width and potential inaccuracies in flux calibration.  

\subsection{KODIAQ at the KOA and \it{igmspec}}\label{koa}

As with DR2, the data products of DR3 are available for community download from the KOA.  As DR3 stems from a different instrument, KOA hosts DR3 as a separate contributed data set from DR2.  Nevertheless, a number of quasars have data in both DR3 and DR2.  Table~\ref{overlap} lists those quasars.  Note that in a few cases, the HIRES DR2 name does not exactly match the ESI DR3 name.  Those cases are highlighted in Table~\ref{overlap}. As with DR2, within DR3 at KOA, users can search for and download individual quasar data, or the full DR3 data set at once.  DR3 will make available all intermediate data reduction products available, grouped by observing run.  Spectra from DR3 will also be ingested into the \textit{igmspec} database\footnote{https://github.com/specdb/igmspec}.

\begin{figure}[ht]
\epsscale{1.25} 
\plotone{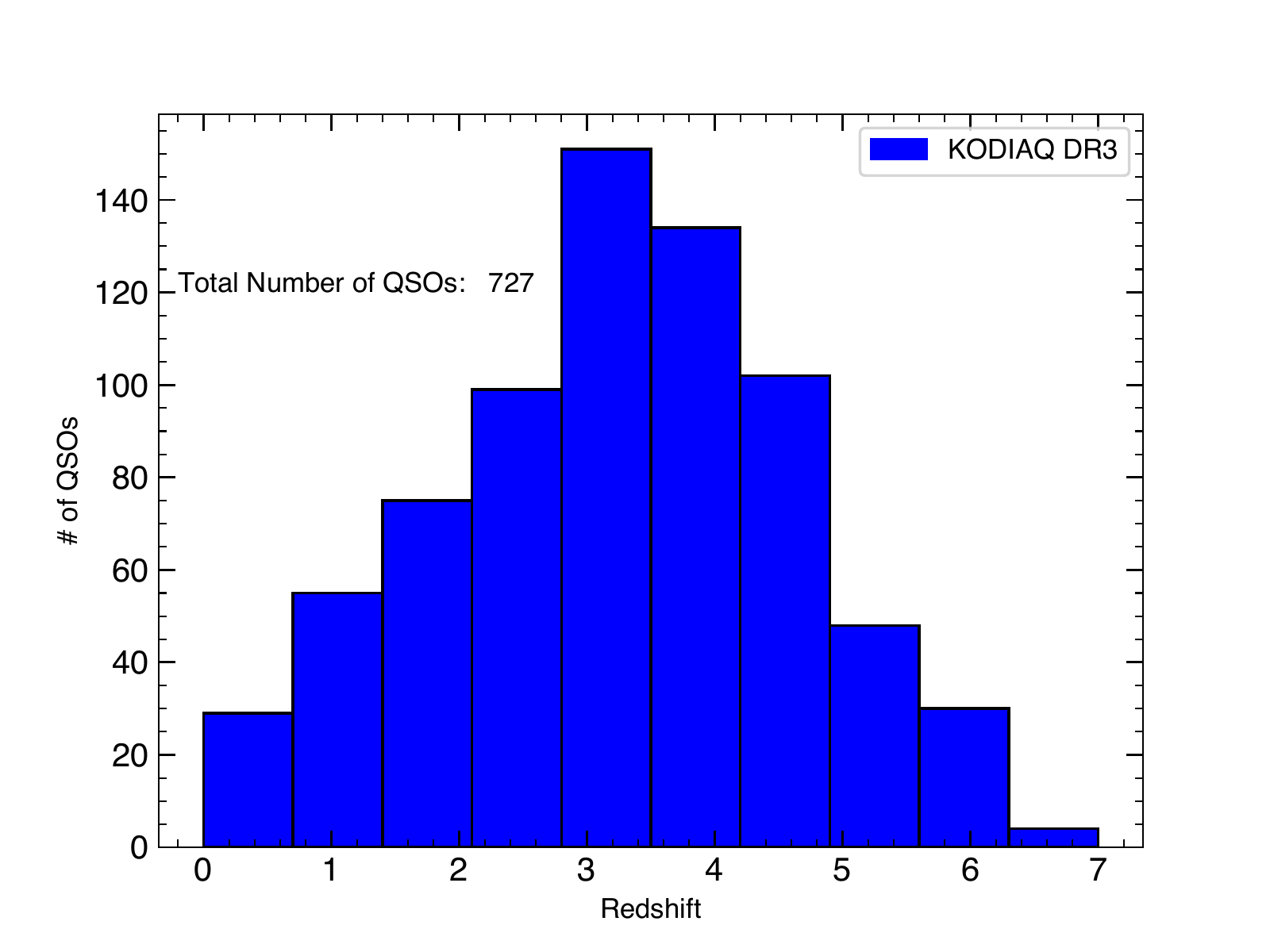}  
\caption{Redshift distribution of KODIAQ DR3 QSOs. \label{fig_zhist}}
\end{figure}

\section{Properties of KODIAQ DR3}\label{dr3}
The KODIAQ DR3 consists of ESI spectra for \nqsos\ quasars.  As in previous data releases, quasars are named according to their J2000 R.A./decl. coordinates as resolved by SIMBAD or NED, unless the quasar is not listed in these databases, in which case we use the position on the sky as provided by the raw data headers.  In Table~\ref{obs_summary} we list the quasar redshift as provided by SIMBAD or NED.  If the quasar is unlisted, a crude estimate is made for the redshift by using the Lyman$-\alpha$ and other emission lines.  The full DR3 is comprised of \ntotspectra\ individual exposures, grouped into \nspectra\ spectral co-adds.  The aggregate exposure time of the DR3 is $\sim2.8$ megaseconds.  

\begin{figure*}[ht]
\epsscale{1} 
\plotone{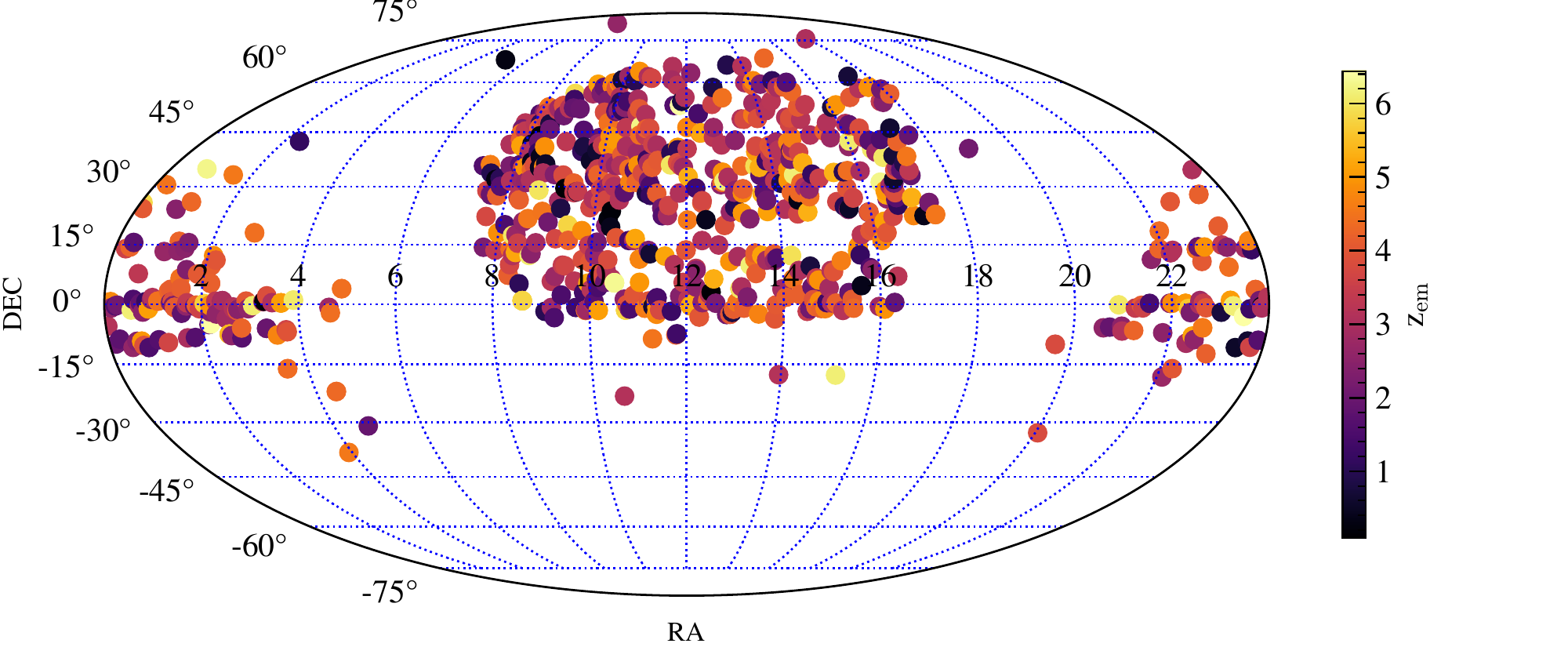}  
\caption{Distribution on the Sky of the KODIAQ DR3 QSOs. \label{fig_skyplot}}
\end{figure*}

\subsection{General Properties}\label{general}
Fig.~\ref{fig_zhist} shows the quasar redshift distribution for the DR3.  The median redshift for the sample is $\bar{z} = 3.21$ and the sample spans the range in redshift of $0.10 < z < 6.44$.  The distribution of DR3 quasars on the sky is shown in Figure \ref{fig_skyplot}.

\begin{figure}[ht]
\epsscale{1.35} 
\plotone{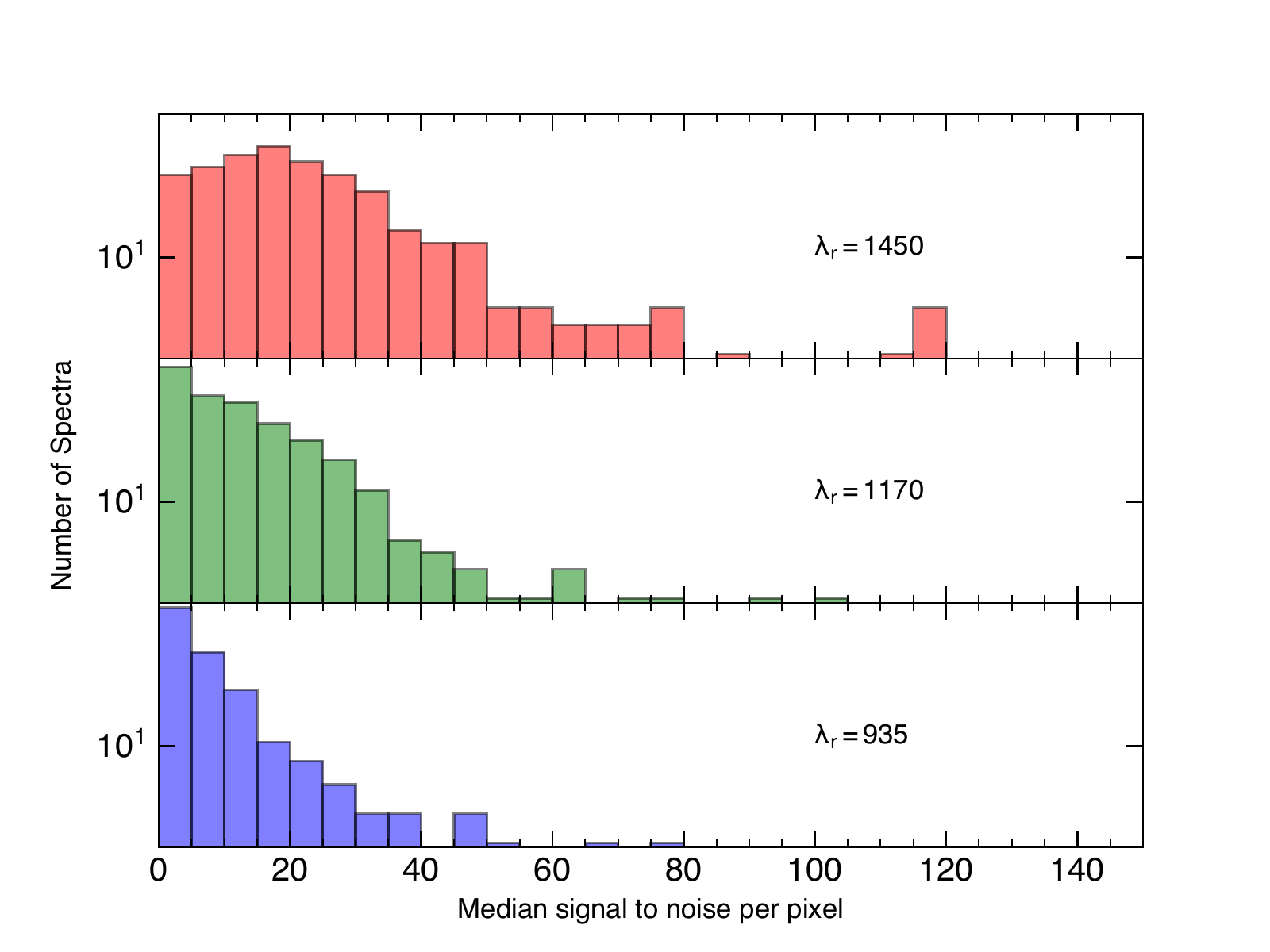}  
\caption{Signal to noise (per pixel) distribution of the spectra in the DR3 sample.\label{fig_snrhist}}
\end{figure}

In Fig.~\ref{fig_snrhist}, we show median signal to noise per pixel, derived from the pipeline optimal extraction values of the flux and 1-$\sigma$ errors, within $\pm 5$ \AA\ for the \nspectra\ in DR3 at 3 rest wavelengths $\lambda_r$ where $\lambda_r = \lambda_{\rm{obs}} / (1 + z_{\rm em})$ and $\lambda_{\rm{obs}}$ is the observed wavelength in the spectrum.  The three values of $\lambda_r$ are chosen as markers for typical wavelengths of interest for IGM/CGM studies, namely $\lambda_r = 1450$\AA\  for studies of heavy element absorption free of Lyman-$\alpha$ forest contamination, $\lambda_r = 1170$ \AA\ for studies of the forest itself, and $\lambda_r = 1450$ \AA\ for determination of the \hi\ column density (\nhi) in Lyman limit systems from the shape of the Lyman break.

\subsection{Cosmological Properties}\label{cosmological}
Given the large number of quasars in DR3, and with a significant fraction of the spectra having moderate to high signal to noise, the sample can lend itself well to a variety of statistical analysis in cosmology and elsewhere.  Here, we highlight a number of statistical properties of DR3 with respect to its ability to be used for studied of neutral hydrogen and heavy element absorption.

\begin{figure}[ht]
\epsscale{1.25} 
\plotone{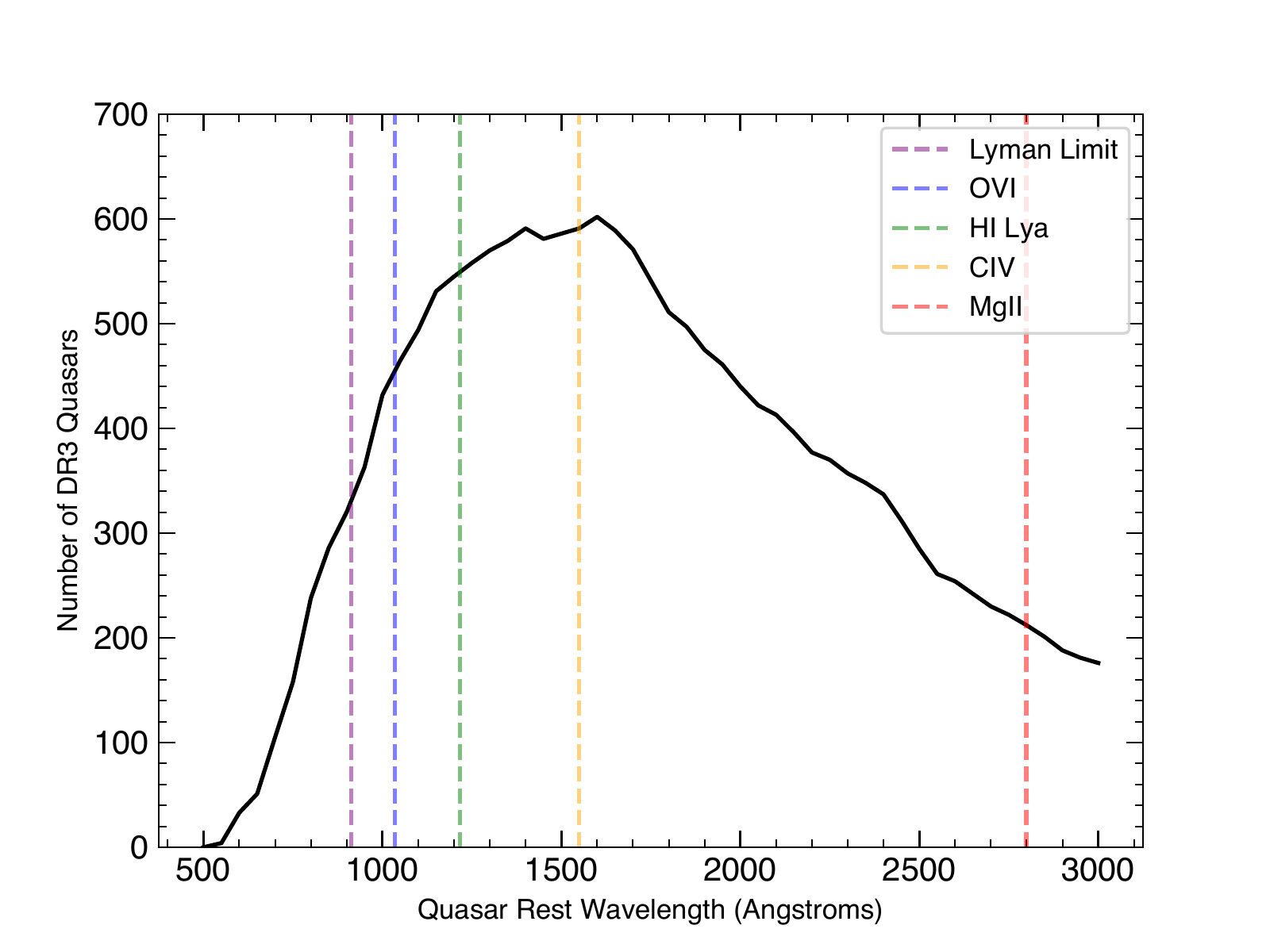}  
\caption{QSO redshift rest wavelength coverage of the DR2 sample. The
  vertical lines correspond to the rest wavelengths of various
  commonly studied ions.\label{fig_resthist}}
\end{figure}

\begin{figure}[ht]
\epsscale{1.25} 
\plotone{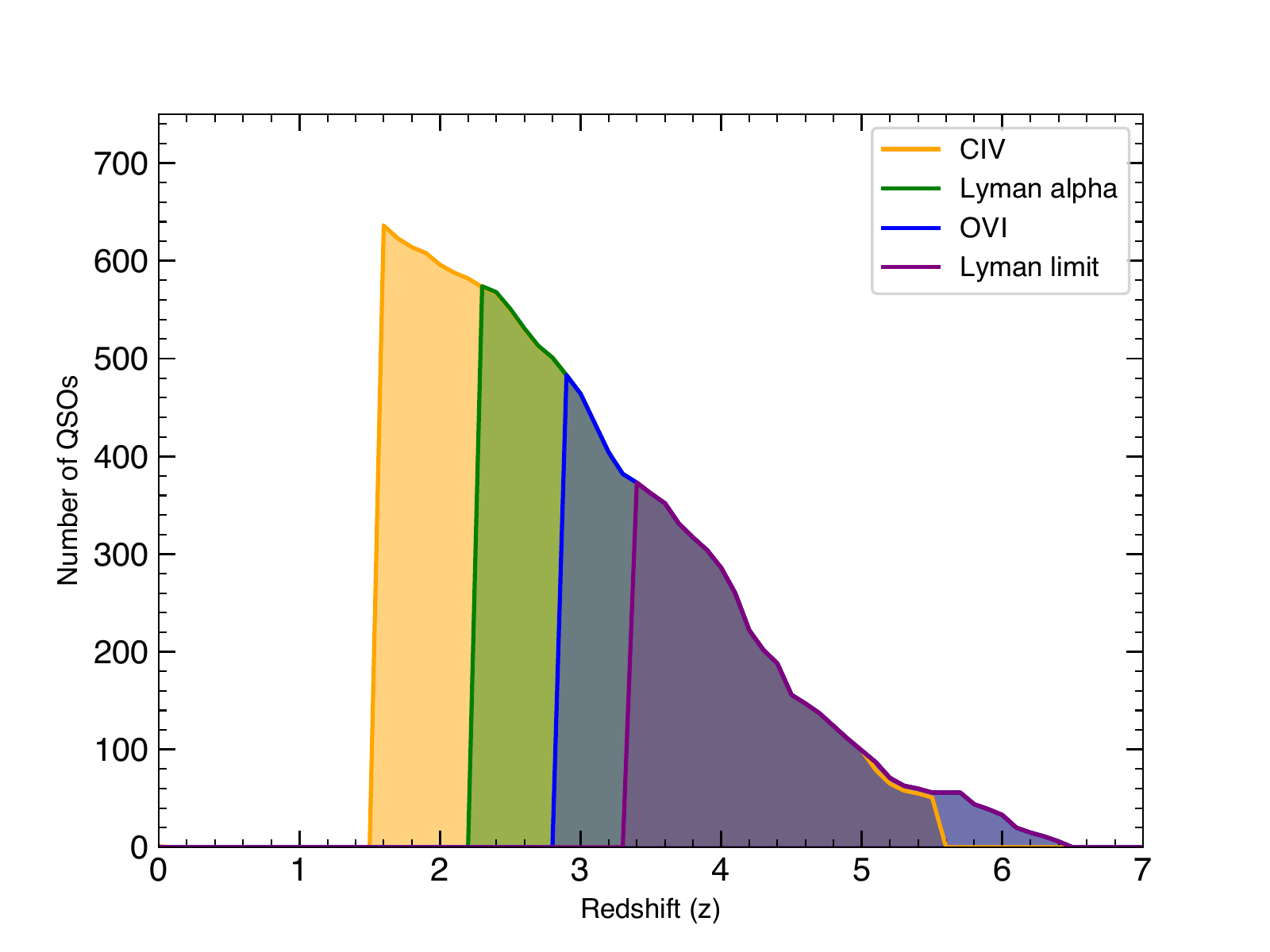}  
\caption{Redshift coverage of various ions in the DR3 sample.\label{fig_ions}}
\end{figure}

\begin{figure*}[ht]
\epsscale{0.9} 
\plotone{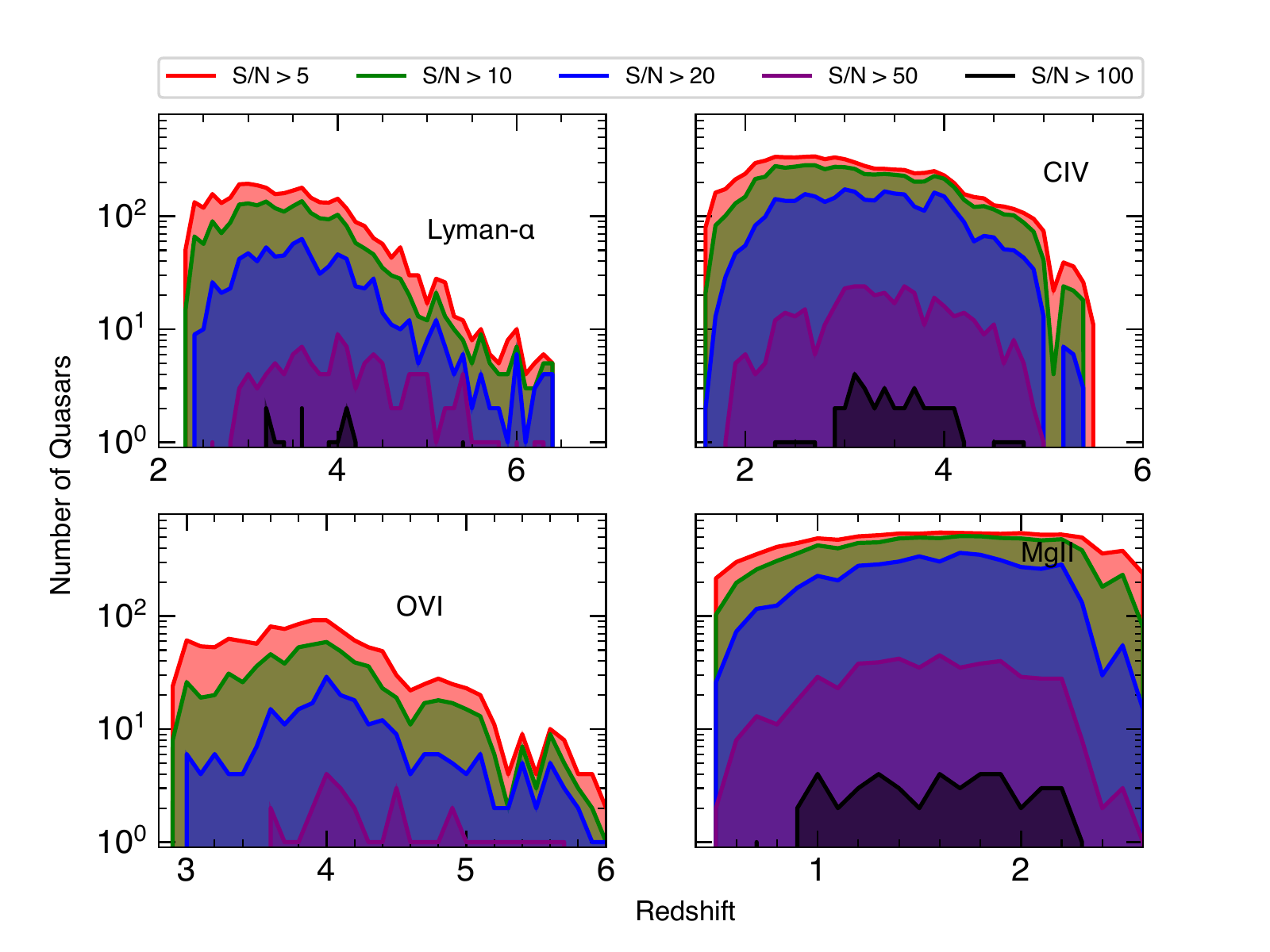}  
\caption{Signal-to-noise  distribution versus redshift for key ions the DR3 sample.\label{fig_snrions}}
\end{figure*}

In Fig.~\ref{fig_resthist} we show the rest wavelength coverage provided by the quasars in DR3 in the quasar rest frame.  Vertical lines denote the rest frame wavelengths of commonly studied lines.  Fig.~\ref{fig_resthist} illustrates that the number of quasars that can be used for statistical studies of \hi\ and heavy element absorption numbers well into the hundreds.  Fig.~\ref{fig_ions} shows this large sample in a different way, namely showing as a function of redshift how many quasars can be used for specific ions. We can further subdivide the DR3 sample by exploring how many quasars provide coverage for specific ions as a function of redshift and signal to noise.  Fig.~\ref{fig_snrions} provides this summary for four ions: \hi\ Lyman-$\alpha$, \civ, \ovi, and \mgii.

\begin{figure*}[ht]
\epsscale{0.9} 
\plotone{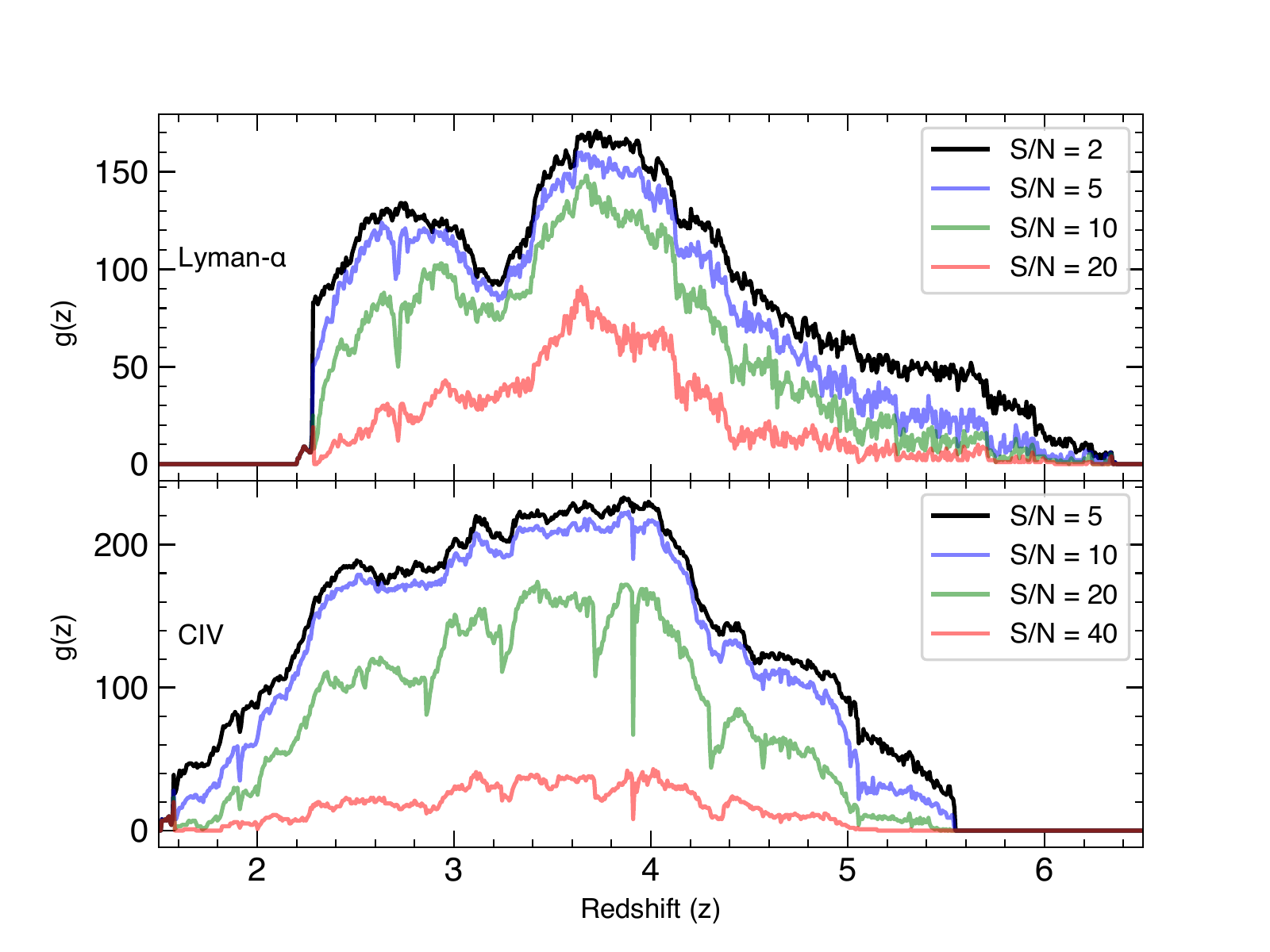}  
\caption{Redshift sensitivity function $g(z)$ for \lya\ and \civ\ in
  the DR3.\label{fig_goz}}
\end{figure*}

Finally, Fig.~\ref{fig_goz} provides the redshift sensitivity function $g(z)$ for the DR3 sample. The function $g(z)$ is calculated in the same manner as in \citet{omeara15} and \citet{omeara17}.  $g(z)$ is presented in Fig.~\ref{fig_goz} for 4 values of the signal-to-noise ratio, and for \hi\ Lyman-$\alpha$ and \civ\ as representative ions.  In all the above metrics, DR3 represents a significant sample for statistical studies of hydrogen and heavy element absorption, even at moderate to high signal-to-noise, which is a requirement for studies of weak absorption.

\section{Summary and Future}\label{future}
We present here a summary of and make publicly available the \nqsos\ quasar sample in the KODIAQ DR3.  As with KODIAQ DR1 and DR2, all spectra and intermediate data reduction products are made available at the KOA.  Unlike DR1 and DR2, the data are roughly uniform in wavelength coverage (and nearly always larger per quasar in wavelength coverage than the HIRES spectra in previous releases), but like DR1 and DR2, they sample a large range of quasar redshift, and provide a wide range of signal-to-noise.

When using data products from DR3, we ask that publications provide the standard KOA acknowledgement, acknowledgment of the original PIs that obtained the data, and the following text:  ``Some or all of the data presented in this work were obtained from the Keck Observatory Database of Ionized Absorbers toward QSOs (KODIAQ), which was funded through NASA ADAP grants NNX10AE84G and NNX16AF52G, along with a citation to this paper.

\section*{Acknowledgements}
Support for this research was made by NASA through the Astrophysics Data Analysis Program (ADAP) grants NNX10AE84G and NNX16AF52G, along with NSF grant award number 1516777. This research has made use of the Keck Observatory Archive (KOA), which is operated by the W. M. Keck Observatory and the NASA Exoplanet Science Institute (NExScI), under contract with the National Aeronautics and Space Administration.   The data presented herein were obtained at the W.M. Keck Observatory, which is operated as a scientific partnership among the California Institute of Technology, the University of California and the National Aeronautics and Space Administration. The Observatory was made possible by the generous financial support of the W.M. Keck Foundation.

The authors wish to recognize and acknowledge the very significant cultural role and reverence that the summit of Maunakea has always had within the indigenous Hawaiian community.  We are most fortunate to have the opportunity to conduct observations from this mountain. The authors wish to recognize and sincerely appreciate the work of the entire WMKO staff over the last two-plus decades, and to the efforts of the team at the NASA Exoplanet Science Institute (NExScI) who are responsible for maintaining the KOA.

\software{XIDL; ESIRedux}
\facilities{KecK(LRIS); Keck(HIRES)}

\bibliographystyle{aasjournal}
\bibliography{ms.bib}

\begin{thebibliography}{}
\expandafter\ifx\csname natexlab\endcsname\relax\def\natexlab#1{#1}\fi
\providecommand{\url}[1]{\href{#1}{#1}}
\providecommand{\dodoi}[1]{doi:~\href{http://doi.org/#1}{\nolinkurl{#1}}}
\providecommand{\doeprint}[1]{\href{http://ascl.net/#1}{\nolinkurl{http://ascl.net/#1}}}
\providecommand{\doarXiv}[1]{\href{https://arxiv.org/abs/#1}{\nolinkurl{https://arxiv.org/abs/#1}}}

\bibitem[{{Bochanski} {et~al.}(2009){Bochanski}, {Hennawi}, {Simcoe},
  {Prochaska}, {West}, {Burgasser}, {Burles}, {Bernstein}, {Williams}, \&
  {Murphy}}]{bochanski09}
{Bochanski}, J.~J., {Hennawi}, J.~F., {Simcoe}, R.~A., {et~al.} 2009, \pasp,
  121, 1409, \dodoi{10.1086/648597}

\bibitem[{{Cooke} {et~al.}(2018){Cooke}, {Pettini}, \& {Steidel}}]{cooke18}
{Cooke}, R.~J., {Pettini}, M., \& {Steidel}, C.~C. 2018, \apj, 855, 102,
  \dodoi{10.3847/1538-4357/aaab53}

\bibitem[{{Hiss} {et~al.}(2019){Hiss}, {Walther}, {O{\~n}orbe}, \&
  {Hennawi}}]{hiss19}
{Hiss}, H., {Walther}, M., {O{\~n}orbe}, J., \& {Hennawi}, J.~F. 2019, \apj,
  876, 71, \dodoi{10.3847/1538-4357/ab1418}

\bibitem[{{Lehner} {et~al.}(2014){Lehner}, {O'Meara}, {Fox}, {Howk},
  {Prochaska}, {Burns}, \& {Armstrong}}]{lehner14}
{Lehner}, N., {O'Meara}, J.~M., {Fox}, A.~J., {et~al.} 2014, \apj, 788, 119,
  \dodoi{10.1088/0004-637X/788/2/119}

\bibitem[{{Murphy}(2018)}]{murphy18}
{Murphy}, M. 2018, {Mtmurphy77/Uves\_Popler: Uves\_Popler: Post-Pipeline
  Echelle Reduction Software}, v1.00,  Zenodo, \dodoi{10.5281/zenodo.1297190}

\bibitem[{{O'Meara} {et~al.}(2017){O'Meara}, {Lehner}, {Howk}, {Prochaska},
  {Fox}, {Peeples}, {Tumlinson}, \& {O'Shea}}]{omeara17}
{O'Meara}, J.~M., {Lehner}, N., {Howk}, J.~C., {et~al.} 2017, \aj, 154, 114,
  \dodoi{10.3847/1538-3881/aa82b8}

\bibitem[{{O'Meara} {et~al.}(2015){O'Meara}, {Lehner}, {Howk}, {Prochaska},
  {Fox}, {Swain}, {Gelino}, {Berriman}, \& {Tran}}]{omeara15}
---. 2015, \aj, 150, 111, \dodoi{10.1088/0004-6256/150/4/111}

\bibitem[{{Prochaska} \& {Wolfe}(2009)}]{prochaska09}
{Prochaska}, J.~X., \& {Wolfe}, A.~M. 2009, \apj, 696, 1543,
  \dodoi{10.1088/0004-637X/696/2/1543}

\bibitem[{{Rafelski} {et~al.}(2012){Rafelski}, {Wolfe}, {Prochaska},
  {Neeleman}, \& {Mendez}}]{rafelski12}
{Rafelski}, M., {Wolfe}, A.~M., {Prochaska}, J.~X., {Neeleman}, M., \&
  {Mendez}, A.~J. 2012, \apj, 755, 89, \dodoi{10.1088/0004-637X/755/2/89}

\bibitem[{{Ross} {et~al.}(2020){Ross}, {Bautista}, {Tojeiro}, {Alam}, {Bailey},
  {Burtin}, {Comparat}, {Dawson}, {de Mattia}, {du Mas des Bourboux},
  {Gil-Mar{\'\i}n}, {Hou}, {Kong}, {Lyke}, {Mohammad}, {Moustakas}, {Mueller},
  {Myers}, {Percival}, {Raichoor}, {Rezaie}, {Seo}, {Smith}, {Tinker},
  {Zarrouk}, {Zhao}, {Zhao}, {Bizyaev}, {Brinkmann}, {Brownstein}, {Carnero
  Rosell}, {Chabanier}, {Choi}, {Chuang}, {Cruz-Gonzalez}, {de la Macorra}, {de
  la Torre}, {Escoffier}, {Fromenteau}, {Higley}, {Jullo}, {Kneib}, {McLane},
  {Mu{\~n}oz-Guti{\'e}rrez}, {Neveux}, {Newman}, {Nitschelm},
  {Palanque-Delabrouille}, {Paviot}, {Pullen}, {Rossi}, {Ruhlmann-Kleider},
  {Schneider}, {Vargas Maga{\~n}a}, {Vivek}, \& {Zhang}}]{sdss20}
{Ross}, A.~J., {Bautista}, J., {Tojeiro}, R., {et~al.} 2020, \mnras,
  \dodoi{10.1093/mnras/staa2416}

\bibitem[{{Sheinis} {et~al.}(2002){Sheinis}, {Bolte}, {Epps}, {Kibrick},
  {Miller}, {Radovan}, {Bigelow}, \& {Sutin}}]{sheinis02}
{Sheinis}, A.~I., {Bolte}, M., {Epps}, H.~W., {et~al.} 2002, \pasp, 114, 851,
  \dodoi{10.1086/341706}

\end{thebibliography}

\startlongtable
\begin{deluxetable*}{lllcccllll} 
\tablewidth{0pc} 
\tablecaption{The KODIAQ DR3 Sample \label{obs_summary}} 
\tabletypesize{\scriptsize} 
\tablehead{ 
\colhead{Object} & 
\colhead{R.A.$^a$} & 
\colhead{Dec.$^a$} & 
\colhead{${z_{\rm em}}^a$} & 
\colhead{Observation} & 
\colhead{PI} & 
\colhead{Program} & 
\colhead{Slit} & 
\colhead{Total Exp,} & 
\colhead{Wavelength} 
\\ 
\colhead{} & \colhead{(J2000)} & \colhead{(J2000)} & \colhead{} & \colhead{Date} & \colhead{} & \colhead{ID$^b$} & \colhead{"} & \colhead{Time (s)}  & \colhead{Coverage (\AA)} 
} 
\startdata
J000211$-$052908  &  00:02:11.7696  &  $-$05:29:08.484  &  2.8  &  Oct 2011  &  Martin  & U047E &  1.00  &  16200.0  &  4000.0--10130.6\\
J000216$-$053007  &  00:02:16.668  &  $-$05:30:07.632  &  3.1592  &  Oct 2011  &  Martin  & U047E &  1.00  &  2400.0  &  4000.0--10130.6\\
J000221+002149  &  00:02:21.1176  &  +00:21:49.356  &  3.0570  &  Dec 2000  &  Becker  & U56E &  0.75  &  900.0  &  4000.0--10134.9\\
J000239+255034  &  00:02:39.3888  &  +25:50:34.8  &  5.82  &  Aug 2005  &  Cowie  & H31aE &  0.75  &  16800.0  &  4000.0--10135.6\\
J000552$-$000655  &  00:05:52.3392  &  $-$00:06:55.8  &  5.85  &  Dec 2002  &  Becker  & XXX &  1.00  &  1200.0  &  4000.0--10134.9\\
\ldots  &  \ldots  &  \ldots  &  \ldots  &  Oct 2010  &  Sargent  & C197E &  0.75  &  15000.0  &  4000.0--10130.9\\
J000738+241724  &  00:07:38.7  &  +24:17:24  &  4.05  &  Sep 2000  &  Djorgovski  & C123E &  1.00  &  3600.0  &  4000.0--9340.51\\
J000749+004119  &  00:07:49.1664  &  +00:41:19.644  &  4.7842  &  Aug 2013  &  Wolfe  & U052E &  0.75  &  3600.0  &  4000.0--10131.9\\
\enddata
\tablecomments{$^a$J2000 coordinates and quasar redshifts are determined by passing the quasar coordinates from the raw data header through SIMBAD or NED and choosing the appropriate match.} 
\tablecomments{$^b$Keck Program ID information taken from the data header.  Data without a Program ID in the header are assigned the value XXX.} 
\end{deluxetable*}

\startlongtable
\begin{deluxetable}{ccc}
\tablecaption{Overlap between DR2 (HIRES) and DR3 (ESI)\label{overlap}}
\tabletypesize{\scriptsize} 
\tablehead{\colhead{ESI DR3 Name} & \colhead{Alternate HIRES DR2 Name} & \colhead{$z_{\rm em}$}}
\startdata
J001602--001225 & J001602--001224 & 2.090 \\
J003501--091817 & & 2.423 \\
J004054--091526 &  & 4.977 \\
J005814+011530 &  & 2.508 \\
J010311+131617 &  & 2.681 \\
J013340+040059 &  & 4.154 \\
J013421+330756 &  & 4.532 \\
J013515--021349 &  & 1.820 \\
\enddata
\end{deluxetable}

\end{document}